\RequirePackage{ifpdf}
\ifpdf 
\documentclass[pdftex]{sigma}
\else
\documentclass{sigma}
\fi

\begin{document}
\allowdisplaybreaks

\renewcommand{\PaperNumber}{038}

\FirstPageHeading

\ShortArticleName{On the Generalized Maxwell--Bloch Equations}

\ArticleName{On the Generalized Maxwell--Bloch Equations}

\Author{Pavle SAKSIDA} \AuthorNameForHeading{P. Saksida}

\Address{Department of Mathematics, Faculty of Mathematics and Physics, \\
University of Ljubljana, Slovenia}

\Email{\href{mailto:Pavle.Saksida@fmf.uni-lj.si}{Pavle.Saksida@fmf.uni-lj.si}}

\ArticleDates{Received December 01, 2005, in f\/inal form March
05, 2006; Published online March 27, 2006}

\newcommand{\li}[1]{\mathfrak #1}
\newcommand{\gl}[1]{\mbox{${\mathfrak #1}$}} 
\newcommand{\Ad}{\mbox{${\rm Ad}$}}
\newcommand{\cgl}[1]{\mbox{$\gl{#1} ^{\mathbb C}$}}   
\newcommand{\hm}{\mbox{${\cal H}$}}    
\newcommand{\hmti}{\mbox{$\widetilde{\hm } $}}  
\newcommand{\grc}{\mbox{$G^{\mathbb C}$}} 
\newcommand{\dmod}{\mbox{${\cal M}_{D^d}$}}
\newcommand{\lam}{\mbox{$\lambda _0$}}
\newcommand{\dcgl}[1]{\mbox{$(\cgl{#1} )^*$}}
\newcommand{\orbc}{\mbox{${\cal O} _{\lam }^{\mathbb C}$}}
\newcommand{\wt}{\mbox{$\widetilde{\tau }$}}
\newcommand{\wg}{\mbox{$\widetilde{G} $}}
\newcommand{\gau}{\mbox{${\cal G} ^{\mathbb C}$}}
\newcommand{\dbara}{\mbox{$\overline{\partial } _A$}}
\newcommand{\dbar}{\mbox{$\overline{\partial }$}}
\newcommand{\dgau}{\mbox{${\cal G} ^{\mathbb C}_D$}}
\newcommand{\kif}{\mbox{${\cal K}$}}
\newcommand{\hit}{\hbox{\bf H}}
\newcommand{\epf}{\hfill   \mbox{$\Box $}}
\newcommand{\dgrc}{\mbox{$G_D^{\mathbb C} $}}
\newcommand{\dg}{\mbox{$deg(D)$}}
\newcommand{\ppar}{\mbox{${\cal M}_{par}$}}
\newcommand{\ddcgl}{\mbox{$\gl{g}_D^{\mathbb C}$}}
\newcommand{\dlam}{\mbox{$\lambda _D $}}
\newcommand{\parmod}{\mbox{$({\cal T}^*{\cal M})_{par}^{\lambda _D}$}}
\newcommand{\prino}{\mbox{${\cal P}{\cal O}_{\lambda _0}^{\mathbb C}$}}
\newcommand{\prinm}{\mbox{${\cal P}M_{\lambda _0}$}}
\newcommand{\pull}{\mbox{${\cal {K(D)}}$}}
\newcommand{\sph}{\mbox{${\mathbb C}{\mathbb P}^1$}}
\newcommand{\rfg}{\mbox{$G_r^{\mathbb C}$}}
\newcommand{\hp}{\mbox{$\gl{h} _{\gl{p}}$}}
\newcommand{\hu}{\mbox{$\gl{h} _{\gl{u}}$}}
\newcommand{\ja}{\mbox{$J\bigl({\cal X}_{\Phi }(z) \bigr)$}}
\newcommand{\rpr}{\mbox{${\mathbb R}{\mathbb P}^n$}}
\newcommand{\ortp}{\mbox{${\cal R}{\cal P}^n$}}
\newcommand{\unip}{\mbox{${\cal C}{\cal P}^n$}}
\newcommand{\cpr}{\mbox{${\mathbb C}{\mathbb P}^n$}}
\newcommand{\qpr}{\mbox{${\mathbb H}{\mathbb P}^n$}}
\newcommand{\simp}{\mbox{${\cal H}{\cal P}^n$}}
\newcommand{\loops}{\mbox{$\Omega G$}}
\newcommand{\rloops}{\mbox{$\widetilde{\Omega }G$}}
\newcommand{\lkif}{\mbox{$\widetilde{{\cal K}}$}}
\newcommand{\ex}{\mbox{${\rm exp}$}}
\newcommand{\tla}{\mbox{$\tau _{\mathfrak{g}}$}}
\newcommand{\tlat}{\mbox{$\tau _{\widetilde{\mathfrak{g}}}$}}
\newcommand{\tlg}{\mbox{$\tau _G$}}
\newcommand{\tlgt}{\mbox{$\tau _{\widetilde{G}}$}}
\newcommand{\sg}{\mbox{$\sigma _G$}}
\newcommand{\sgt}{\mbox{$\sigma _{\widetilde{G} }$}}
\newcommand{\sa}{\mbox{$\sigma _{\mathfrak{g} }$}}
\newcommand{\sat}{\mbox{$\sigma _{\widetilde{\mathfrak{g} }}$}}
\newcommand{\spr}{\mbox{${\mathbb S}^n$}}
\newcommand{\hyp}{\mbox{${\mathbb H}^n$}}
\newcommand{\cf}{\mbox{${\mathbb C}$}}
\newcommand{\rf}{\mbox{${\mathbb R}$}}
\newcommand{\rp}{\mbox{${\mathbb R}{\mathbb P}$}}
\newcommand{\cp}{\mbox{${\mathbb C}{\mathbb P}$}}
\newcommand{\rl}{\mbox{${\widetilde{L} \grc }$}}
\newcommand{\lrl}{\mbox{${\widetilde{L} \cgl{g} }$}}
\newcommand{\lel}[1]{\mbox{${\widetilde{#1}}$}}
\newcommand{\rlrl}{\mbox{${\widetilde{L} \gl{g} }$}}
\newcommand{\rrl}{\mbox{${\widetilde{L} G}$}}
\newcommand{\lb}{\mbox{{\bf [ }}}
\newcommand{\rb}{\mbox{{\bf \   ]}}}
\newcommand{\hrp}{\mbox{${H_{{\mathbb R{\mathbb P}}^n}}$}}
\newcommand{\hcp}{\mbox{${H_{{\mathbb C{\mathbb P}}^n}}$}}
\newcommand{\ver}{\mbox{${\rm Vert}_z$}}
\newcommand{\hor}{\mbox{${\rm Hor}_z$}}
\newcommand{\dd}{\mbox{${\rm d}$}}
\newcommand{\su}{\mbox{$\mathfrak{s} \mathfrak{u}$}}
\newcommand{\sll}{\mbox{$\mathfrak{s} \mathfrak{l}$}}
\newcommand{\rd}{\mbox{${\cal R}(x)$}}
\newcommand{\leb}{\mbox{$\langle \! \langle$}}
\newcommand{\rib}{\mbox{$\rangle \! \rangle$}}
\newcommand{\orb}{\mbox{${\cal O}_{\tau}$}}

\Abstract{A new Hamiltonian structure of the Maxwell--Bloch
equations is described. In this setting the Maxwell--Bloch
equations appear as a member of a family of generalized
Maxwell--Bloch systems. The family is parameterized by compact
semi-simple Lie groups, the original Maxwell--Bloch system being
the member corresponding to $SU(2)$. The Hamiltonian structure is
then used in the construction of a new family of symmetries and
the associated conserved quantities of the Maxwell--Bloch
equations.}


\Classification{37K05; 35Q60; 37K30; 35Q58; 53D20}

\section{Introduction}

Maxwell--Bloch equations are a system of partial dif\/ferential
equations which plays a prominent role in the f\/ield of
non-linear optics. This system models the resonant interaction
between light and an optically active medium consisting of
two-level atoms. The quantities f\/iguring in this system are the
complex valued functions $E(t, x), P(t, x) \colon \rf ^2 \to \cf$
and a real valued function $D(t, x) \colon \rf ^2 \to \rf$ of two
independent variables, time $t$ and one spatial variable $x$.  The
function $E(t, x)$ is the slowly varying envelope of the electric
f\/ield, $P(t,x)$ is the polarization of the medium and $D(t, x)$
is the population inversion. Here we will consider the
Maxwell--Bloch equations without pumping and in the sharp-line
limit, that is, without inhomogeneous broadening:
\begin{gather}
E_t + c E_x = P , \qquad P_t = E D -  \beta P, \qquad D_t = -
\frac{1}{2} (\overline{E} P + E \overline{P}). \label{mbe}
\end{gather}
The constant $c$ above is the speed of light in the medium and
$\beta$ represents the longitudinal relaxation time of the medium.
The transverse relaxation time and the losses of the electric
f\/ield are assumed to be equal to zero.

We will show that the Maxwell--Bloch equations describe a
continuous chain of interacting C.~Neumann oscillators on the
three-sphere. The interactions between the neighbouring
oscillators are of magnetic type, which means that the
acceleration of any given oscillator depends on the velocity of
its neighbours and not on their position. More explicitly, we will
show that the system~(\ref{mbe}) is equivalent (modulo a certain
constraint) to the second-order partial dif\/ferential equation
\begin{gather}
\big(g_tg^{- 1}\big)_t + c  \big(g_t g^{-1}\big)_x = [ \sigma,
\Ad_g(\tau (x))], \label{MBN}
\end{gather}
where $g(t, x) \colon \rf \times \rf \to  SU(2)$ is a Lie group
valued function of two variables, $\sigma \in \su (2)$ is a
constant and $\tau(x) \colon \rf \to \su (2)$ an arbitrary path in
the Lie algebra. We shall see that this equation is the equation
of motion of the above-mentioned chain of C.~Neumann oscillators.

If we replace the group $SU(2)$ by an arbitrary Lie group $G$, the
equation (\ref{MBN}) still makes sense. We shall call the equation
(\ref{MBN}), with $g(t, x) \colon \rf \times \rf \to G$ and
$\sigma, \tau (x) \in \gl {g} = {\rm Lie}(G)$, the  generalized
Maxwell--Bloch equation. We will show that in the case when $G$ is
a compact semi-simple Lie group, the equation (\ref{MBN}) is again
the equation of motion for a continuous chain of oscillators. In
this case, the conf\/iguration space of the relevant model
oscillator will be the Lie group $G$. These oscillators belong to
a class of  well-known integrable systems described by Reyman and
Semenov-Tian-Shansky in \cite{ReSe1} and \cite{ReSe2}, and later
in dif\/ferent contexts by other authors, see e.g.~\cite{Sa3}.

The rewriting (\ref{MBN}) of the Maxwell--Bloch equations proves
to be useful in several ways. In this paper, we shall concentrate
on two features. First we shall construct and describe a new
Hamiltonian structure for the Maxwell--Bloch equations and their
generalizations. This will then enable us to f\/ind an
inf\/inite-dimensional group of symmetries and corresponding
inf\/inite-dimensional class of conserved quantities of our
equation.

We shall limit ourselves to the spatially periodic case
\[
g(t, x + 2 \pi) = g(t, x), \qquad \tau (x + 2 \pi) = \tau (x),
\]
and we shall assume that the group $G$ is compact and semi-simple.
We will show that (\ref{MBN}) is the equation of motion for the
Hamiltonian system $(T^*LG, \omega_c + c   \omega_m, H_{mb})$.
Here the conf\/iguration space $LG = \{ g(x) \colon S^1 \to G\}$
is the loop group over $G$. The Hamiltonian $H_{mb} \colon T^*LG
\to \rf$ is given by the formula
\[
H_{mb} (g, p_g) = \int _{S^1} \left(\frac{1}{2} \| p_g(x) \|^2  +
K(\sigma,  \Ad_{g(x)} \tau (x)  ) \right) \dd x,
\]
where $K(-, -)$ is the Killing form on $\gl {g}$. The symplectic
structure $\omega _c + c  \omega _m$ is the canonical structure
$\omega _c$ on the cotangent bundle $T^*G$, perturbed by the so
called magnetic term $ c   \omega _m$. The term $\omega _m$ is the
pull-back $\pi^*(\widetilde{\omega}_m)$, via the natural
projection $\pi \colon T^*LG \to LG$, of a right-invariant
dif\/ferential 2-form $\widetilde{\omega}_m$ on $LG$. The value of
$\widetilde{\omega}_m$ at the identity is given by
\[
(\widetilde{\omega} _m)_e (\xi, \eta ) = \int _{S^1} K( \xi '(x),
 \eta(x) ) \, \dd x,
\qquad \xi(x),  \eta(x) \in L \gl{g} = T_e LG(2).
\]
We stress that in the case when $G = SU(2)$, the above system
provides a {\it new} Hamiltonian structure for the Maxwell--Bloch
equations (\ref{mbe}).

We see that the symplectic structure of our Hamiltonian system is
not canonical. It is well-known that the perturbations of the
canonical symplectic forms are responsible for the forces of the
magnetic type, see e.g.~\cite{No,Mar1,Mar2}. The model example of
such a situation is the motion of an electrically charged particle
in a magnetic f\/ield. The Lorentz force can be encoded  as
a~perturbation of the canonical symplectic form on $T^*\rf^3$. One
can geometrize the Lorentz-type force by adding another (circular)
degree of freedom to the conf\/iguration space. On the suitably
extended phase space $T^*M$ the symplectic form will be canonical.
One can then easily f\/ind such a metric on  $M$ that the
geodesics on $M$ will project down to the trajectories of a
particle in $N$ under the inf\/luence of our magnetic-type force.
This procedure is known by the name of the  Kaluza--Klein theory.
Whenever the magnetic perturbation $\omega_m$ is exact, the
Kaluza--Klein extended space is simply $M = N \times U(1)$. If
$\omega_m$ is not exact, then the extended space  exists under the
condition that $\omega_m$ has a certain integrality property. In
this case, the extended space is a non-trivial $U(1)$-bundle whose
Chern class is equal to the de Rham class of $\omega _m$. In the
case of the generalized Maxwell--Bloch equations the form
$\widetilde{\omega}_m \in \Omega^2 LG$ is {\it not} exact. The
extended conf\/iguration space  is therefore a non-trivial
$U(1)$-principal bundle over $LG$.  It is actually  precisely the
central extension $\widetilde{L}G$  of the loop group $LG$. A
detailed exposition of the Kaluza--Klein description of the
Maxwell--Bloch equations can be found in~\cite{Sa}. The geometric
prerequisites, needed for the construction of non-trivial extended
spaces, are given in~\cite{Ko}. Here, we would only like to
mention the following interesting fact: In the Kaluza--Klein
description of the electron moving in a magnetic f\/ield, the
moment, conjugate to the additional circular degree of freedom, is
the charge of the electron. Therefore, also in other situations,
this moment is called charge. It is interesting to note that in
the case of the Maxwell--Bloch equations the charge is precisely
the speed of light in the medium.

As we already mentioned, the important merit of the equation
(\ref{MBN}) and the corresponding Hamiltonian system $(T^*LG,
\omega_c + c   \omega _m, H_{mb})$ is the fact that they enable us
to f\/ind an inf\/inite-dimensional family of symmetries and the
corresponding conserved quantities of the (generalized)
Maxwell--Bloch equation. We will show in the
Section~\ref{symmetries} that an Abelian loop group $LT$, where
$T$ is a certain maximal torus in~$G$, acts in a Hamiltonian way
on our system. The  corresponding conserved quantities will be
constructed as the components of the moment map associated to the
action of $LT$. Thus, we shall obtain a new family of conserved
quantities.

The rewriting (\ref{MBN}) of the Maxwell--Bloch equations appeared
for the f\/irst time in the papers \cite{Q-HPS1} and \cite{Q-HPS2}
by Q-Han Park and H.J. Shin. Without previous knowledge of these
papers, the author rediscovered this rewriting and used it in the
paper~\cite{Sa}. In \cite{Q-HPS1} and \cite{Q-HPS2} the equation
(\ref{MBN}) is viewed as an equation of f\/ield theory, while
in~\cite{Sa} it is treated as a continuous chain of C.~Neumann
oscillators. With respect to the generalized Maxwell--Bloch
equation it is interesting to note the following. The authors of
\cite{Q-HPS2} show that certain cases of the generalized
Maxwell--Bloch equation (\ref{MBN}) with $SU(2)$ replaced, say, by
$SU(3)$ or $SU(4)$ are actually physically meaningful. They
describe various instances of degenerate and non-degenerate
interaction of light with two and three level optical media. The
correct choice of $\tau(x)$ is crucial here.

\section[Generalized Maxwell-Bloch equation]{Generalized Maxwell--Bloch equation}

First we will show that the equations (\ref{mbe}) and (\ref{MBN})
are indeed equivalent. Let us introduce the matrix valued
functions $\rho(t, x) \colon \rf \times \rf \to \su (2)$ and $F(t,
x)  \colon \rf \times \rf \to \su (2)$ by the formulae
\begin{gather}
\rho (t, x)   =  \begin{pmatrix} i D(t, x) & i P(t, x)  \\
                       -\overline{i P}(t, x) & - i D(t, x) \end{pmatrix}
                      , \qquad
 F(t, x) = \frac{1}{2}\begin{pmatrix} i \beta & E(t, x) \\
              - \overline{E}(t, x) & - i \beta \end{pmatrix}.
\label{EPDtoRF}
\end{gather}
In terms of $F$ and $\rho$ the Maxwell--Bloch equations become
\begin{gather}
\rho _t   =   [\rho, F],   \qquad
 F_t +  c   F_x   =  [\rho, \sigma],
\label{MBE2}
\end{gather}
where
\[
\sigma = \frac{1}{2} \begin{pmatrix} i & 0 \\
                              0  & - i \end{pmatrix}
\]
is the f\/irst Pauli matrix multiplied by $i$. The f\/irst of the
above equations is a Lax equation. Its general solution is of the
form
\begin{gather}
\rho (t, x) = \Ad _{g(t, x)} (\tau (x)),  \qquad F(t, x) = -
g_t(t, x)\cdot g^{-1}(t, x). \label{RFtoG}
\end{gather}
Here $g(t, x) \colon \rf \times \rf \to SU(2)$ is a Lie group
valued function and $\tau(x) \colon \rf \to \su (2)$ takes values
in the Lie algebra $\su (2)$. If we insert (\ref{RFtoG}) into the
second equation of~(\ref{MBE2}), we indeed get the second order
equation
\begin{gather}
\big(g_tg^{- 1}\big)_t + c   \big(g_t g^{-1}\big)_x = [ \sigma,
\Ad_g(\tau (x))]. \label{MBN1}
\end{gather}
We note that the diagonal terms of the matrix $F(t, x)$ are
constant. Thus we see that Maxwell--Bloch equations are equivalent
to the equation (\ref{MBN1}) together with the constraint
\[
\langle g_t g^{-1}, \sigma \rangle = {\rm const} = - \beta.
\]

From the point of view of the original physical interpretation, it would be better to introduce
the variables $\rho$ and $F$ in the formula (\ref{EPDtoRF}) as
Hermitian rather than skew-Hermitian matrices. This would be
achieved by multiplying the matrices by $- i$. Then $\rho$ would
indeed have the proper form of the projection on the quantum
mechanical wave-function. We have adopted the skew-Hermitian form
because it is more convenient from the mathematical point of view
on the one hand, and because it is better suited to our mechanical
interpretation of the Maxwell--Bloch equations on the other.
Namely, being skew-Hermitian, the matrices $\rho$ and $F$ are the
proper  elements of the Lie algebra $\su (2)$ and this will
facilitate our reasoning and calculations below.

From now on we shall denote  the Killing form by $\langle - , - \rangle$.
In the equation (\ref{MBN1}) we can replace the unknown function
$g(t, x) \colon \rf \times \rf \to SU(2)$ by the function $g(t, x)
\colon \rf \times \rf \to  G$ which takes values in an arbitrary
Lie group $G$, if only we replace $\sigma \in \su (2)$ and $\tau
(x) \in \su (2)$ by an element $\sigma \in \gl{g}$ and a curve
$\tau (x) \colon \rf \to \gl{g}$, where $\gl{g}$ is the Lie
algebra of $G$.

We can simplify the equation (\ref{MBN1}) to some  extent. First
we note the following. Let $g(t, x) \colon \rf \times \rf \to G$
be a solution of (\ref{MBN1}). Then it is easy to check that for
every $h(x) \colon \rf \to G$ the function $ f(t, x) = g(t, x)
h(x) \colon \rf \times \rf \to G $ is a solution of the equation
\[
\big(f_t f^{-1})_t + c   \big(f_t f^{-1}\big)_x = \big[ \sigma,
\Ad _f \bigl(h(x) \cdot \tau (x) \cdot h^{-1}(x)\bigr) \big].
\]
Therefore, we can assume that $\tau (x)\colon \rf \to \gl{t}
\subset \gl{g}$ is a map which takes values in a chosen maximal
toroidal subalgebra $\gl{t} \subset \gl{g}$.

We obtain an important equation, if we perform the following
reduction. Consider the equation (\ref{MBN1}) for the case when $G
= SU(2)$, but let the unknown function $g(t, x)$ be constrained to
take values only in a subgroup $U(1) \subset SU(2)$:
\[
g(t, x) = \begin{pmatrix} \cos{\phi(t, x)} & \sin{\phi (t, x)} \\
                          - \sin{\phi (t, x)} & \cos{\phi(t, x)} \end{pmatrix}
   \colon \rf \times \rf \to U(1) \subset SU(2).
 \]
 Let $\sigma $ be the Pauli matrix $\sigma = \frac{1}{2}  \,{\rm diag}\,(i, - i)$ and let
 $\tau (x) \equiv\sigma$. Then a calculation shows that the equation
 (\ref{MBN1}) gives
 \[
 \phi _{tt} +  c   \phi _{tx}  = \sin {\phi}.
 \]
 This equation is essentially the sine-Gordon equation.

\section{Hamiltonian structure}

Let us consider the spatially constant solutions $g(t) \colon \rf
\to G$ of the equation (\ref{MBN1}). These solutions actually
solve the  equation
\begin{gather}
\big(g_tg^{-1}\big)_t = [\sigma, \Ad_g(\tau)]. \label{ode}
\end{gather}
The following proposition describes the Hamiltonian nature of the
above ordinary dif\/ferential equation.

\begin{proposition}\label{prop2}
The equation \eqref{ode} is the equation of motion of the  system
$(T^*G, \omega_c, H)$, where $\omega_c $ is the canonical
symplectic form on the cotangent bundle $T^*G$ and the Hamiltonian
function $H \colon T^*G \to \rf$ is given by
\begin{gather}
H(g, p_g) = \frac{1}{2} \| p_g \| ^2 + \langle \sigma, \Ad_g(\tau)
\rangle. \label{cnHam}
\end{gather}
\end{proposition}

\begin{proof}
Let us trivialize the cotangent bundle $T^*G$ over the Lie group
$G$ by means of the right translations, $T^*G \cong G \times
\gl{g}$. With this trivialization in mind we shall denote the
elements in the tangent spaces $T_{(g, p_g)}(T^*G) \cong \gl{g}
\times \gl{g}^*$ by $(X_b, X_{ct})$. The canonical symplectic form
$\omega_c $ on the cotangent bundle $T^*G$ over the Lie group $G$
can be given by the formula
\begin{gather}
(\omega_c)_{(g, p_g)}((X_b, X_{ct}), (Y_b, Y_{ct})) = - \langle
X_{ct}, Y_b \rangle + \langle Y_{ct}, X_b \rangle + \langle p_g,
[X_b, Y_b] \rangle. \label{ctg}
\end{gather}
Above $\langle - , - \rangle$ denotes the pairing between the
elements of $\gl{g}$ and those of $\gl{g}^*$. For the proof of
this formula see \cite{AbMa}.

A solution of a Hamiltonian system is an integral curve of the
Hamiltonian vector f\/ield $X_H$ and thus is given by the relation
$\dd H = \omega (X_H, -)$, where $\omega$ is the symplectic form.
For the Hamiltonian given by (\ref{cnHam}) we have
\begin{gather}
\langle \dd H, (\delta _b, \delta _{ct}) \rangle = - \langle
[\sigma, \Ad_g(\tau)]^{\flat}, \delta_b \rangle + \langle \delta
_{ct}, p_g^{\sharp} \rangle , \label{dif}
\end{gather}
where $\flat \colon \gl{g} \to \gl{g} ^*$ and $\sharp \colon
\gl{g}^* \to \gl{g}$ are def\/ined by $\alpha ^{\flat} = \langle
\alpha, - \rangle$ and $\beta = \langle \beta^{\sharp}, -
\rangle$. Let us denote $X_H = (X_b, X_{ct})$. Then we have
\begin{gather}
(\omega _c)_{(g, p_g)} ((X_b, X_{ct}), ( \delta _b, \delta _{ct}))
= - \langle X_{ct}, \delta _b \rangle + \langle  \delta _{ct}, X_b
\rangle +
\langle p_g, [X_b, \delta_b ] \rangle \nonumber  \\
\phantom{(\omega _c)_{(g, p_g)} ((X_b, X_{ct}), ( \delta _b,
\delta _{ct}))}{}
 =\langle - X_{ct} - \{ X_b, p_g\}, \delta _b \rangle +  \langle \delta _{ct}, X_b \rangle ,
\label{plug}
\end{gather}
where $\{a, \alpha \}$ denotes the ${\rm ad}^*$-action of $a\in
\gl{g}$ on $\alpha \in \gl{g}^*$. The above equations now yield
\[
p_g^{\sharp} = X_b, \qquad [\sigma,\Ad_g(\tau)]^{\flat} =  X_{ct}
+ \{X_b, p_g\},
\]
and therefore
\[
X_b = p_g^{\sharp}, \qquad  X_{ct} = [\sigma,
\Ad_g(\tau)]^{\flat}.
\]
Let now $\gamma(t) = (g(t), p_g(t)) \colon \rf \to T^*G$ be a
curve expressed in the right trivialization and let $\dot{\gamma}
= (g_tg^{-1}, (p_g)_t)$ be its tangent at the point $(g(t),
p_g(t))$. The  above equations tell us that $\gamma (t)$~is an
integral curve of the Hamiltonian vector f\/ield $X_H$, if and
only if it satisf\/ies the equation
\begin{gather*}
(g_tg^{- 1} )_t = [\sigma, \Ad_g(\tau)]. \tag*{\qed}
\end{gather*}\renewcommand{\qed}{}
\end{proof}

In the case when $G = SU(2)$, the system $(T^*SU(2), \omega_c, H)$
is the C.~Neumann system which describes the motion of a particle
on $S^3 = SU(2)$ under the inf\/luence of a quadratic potential.
The quadratic form def\/ining the potential has two double
eigenvalues. To see this we only have to calculate explicitly the
potential $\langle \sigma,\Ad_g(\tau) \rangle$ in the case when $g
\in SU(2)$. The elements of $SU(2)$ are matrices of the form
\[
g = \begin{pmatrix} g_1 + i g_2 & g_3 + i g_4 \\
              - g_3 + i g_4 & g_1 - i g_2 \end{pmatrix}
              , \qquad \det{(g)}=  \sum _{i = 1}^4 g_i^2= 1 .\]
 If we take
 \[
 \tau = \begin{pmatrix} i a & b + i c \\
                 - b + i c & - ia \end{pmatrix} \qquad {\rm and}
                 \qquad \sigma =
                 \begin{pmatrix}i & 0 \\
                                0 & - i \end{pmatrix},
\]
we get
\begin{gather*}
2 \langle \sigma , \Ad_g(\tau)\rangle = - {\rm Tr}\,\big (\sigma g \tau g^{-1}\big)\\
\phantom{2 \langle \sigma , \Ad_g(\tau)\rangle}{} =  2 a (g_1^2 +
g_2^2 - g_3^2 - g_4^2) + 4b (- g_1 g_4 + g_2 g_3) + 4c (g_1 g_3 +
g_2g_4).
\end{gather*}
The matrix of the above quadratic form has indeed two double
eigenvalues $ \lambda = 2 \| \tau\| $ and $ \mu = - \lambda $.
Thus the system $(T^*G, \omega_c, H)$, where the Hamiltonian is
given by (\ref{cnHam}), can be considered as  a~generalized
C.~Neumann oscillator. Its conf\/iguration space is the  Lie group
$G$ instead of the sphere.

The fact that the quadratic form of the potential of $(T^*SU(2),
\omega_c, H)$ has two double eigenvalues means that the system
has two circular symmetries. Moreover, the system $(T^*G,
\omega_c, H)$, where $G$ is compact semi-simple,  has generically
two toroidal symmetries. Inspection of the Hamiltonian
(\ref{cnHam}) immediately shows that our system is preserved under
the right action $\rho^r_t(g) = g \cdot t$ of the maximal torus
$T_{\tau} = \exp{\gl{t}_{\tau}}$ and under the left action
$\rho^l_t(g) = t \cdot g$ of the torus $T_{\sigma} =
\exp{\gl{t}_{\sigma}}$. Here $\gl{t}_{\tau}$ and $\gl{t}_{\sigma}$
denote the maximal toroidal subalgebras in $\gl{g}$ which contain
the elements $\tau$ and $\sigma$, respectively. We have taken into
the account the fact that the canonical symplectic form $\omega _c
$ on $T^*G$ is left and right-invariant. The symmetries $\rho^r$
and $\rho^l$ will enable us to construct in Section
\ref{symmetries}  the   symmetries of the generalized
Maxwell--Bloch equation.

Now we shall return to the partial dif\/ferential equation
(\ref{MBN1}). We shall concentrate on the spatially periodic case,
which means that we shall stipulate
\[
g(t, x + 2 \pi) = g(t, x), \qquad \tau (x + 2 \pi ) = \tau (x).
\]
In a somewhat more graphic way, the equation (\ref{MBN1}) can be
written as
\[
\big(g_tg^{-1}\big)_t(t, x) =  - \frac{c  }{\epsilon} \bigl(
g_tg^{-1} (t, x - \epsilon) - g_tg^{-1}(t, x + \epsilon)\bigr)
\big|_{\epsilon \to 0} + [\sigma, \Ad_{g(t, x)} (\tau (x))].
\]
Consider the function $g(t, x_0)$ at a f\/ixed value $x_0$ of the
spatial coordinate. We can think of~$g(t, x_0)$ as of the position
of the generalized C.~Neumann oscillator moving in $G$. Its
acceleration depends on the potential $[\sigma,\Ad _g(\tau)]$ and
on the velocities $(g_tg^{-1})(t, x_0 \pm \epsilon)$ of the
neighbouring oscillators. It is thus natural to try to understand
the equation (\ref{MBN1}) as the equation of motion for a
continuous chain of the generalized  C.~Neumann oscillators. The
oscillators interact among themselves in a magnetic way. By this
we mean that the acceleration of every given oscillator depends on
the velocity and not on the position of its neighbours.

Let us now construct the Hamiltonian structure of the generalized
Maxwell--Bloch equation which will correspond to the above
interpretation. The obvious candidate for the conf\/iguration
space of the continuous chain is the set of maps $g(x) \colon S^1
\to G$, or, in other words, the loop group $LG$. The phase space
is therefore the cotangent bundle $T^*LG$. The natural choice for
the Hamiltonian is the total energy of all the oscillators:
\begin{gather}
H_{mb}\bigl(g(x), p_g(x)\bigr) = \int _{S^1} \left( \frac{1}{2} \|
p_g(x)\|^2 + \langle \sigma, \Ad_{g(x)}(\tau (x))\rangle \right)
\dd x. \label{MBEH}
\end{gather}
But the Hamiltonian system $(T^*LG, \omega_c, H_{mb})$ does not
correspond to the generalized Maxwell--Bloch
equation~(\ref{MBN1}). It is easily seen that the equation of
motion of this system is simply the equation $(g_tg)_t(t, x) =
[\sigma, \Ad_{g(t, x)}(\tau(x))]$ which describes the system of
decoupled generalized C.~Neumann oscillators. Therefore we have to
modify this system in such a way that the magnetic-type
interactions will be taken into account. We shall achieve this by
perturbing the canonical symplectic form $\omega_c$ by an
additional term.

Let $\Omega $ be the cocycle on the loop algebra $L\gl{g}$
def\/ining the central extension $\widetilde{L} \gl{g} = L \gl{g}
\times \rf$.  Recall that $\Omega$ is given by
\[
\Omega (\xi (x), \eta(x)) = - \int _{S^1} \langle \xi ' (x), \eta
(x) \rangle, \qquad \xi (x), \eta (x) \in L\gl{g}.
\]
Let us denote by $\widetilde{\omega}_m$ the right-invariant 2-form
on the loop  group $LG$ whose value at the identity is equal to
$\Omega$, that is $(\widetilde{\omega}_m)_e = \Omega$. Finally let
$\pi \colon T^* LG \to LG$ be the natural projection and let the
2-form $\omega _m$ on $T^*LG$ be given as the pull-back $\omega_m
= \pi^*(\widetilde{\omega}_m)$. We have the following theorem.
\begin{theorem}
Let the Hamiltonian $H_{mb}$ in the Hamiltonian system $(T^*LG,
\omega_c + c   \omega _m, H_{mb})$ be given by \eqref{MBEH}, and
let $\omega_m$ be of the form described above. Then the equation
of motion of this Hamiltonian system is the generalized
Maxwell--Bloch equation~\eqref{MBN1}.
\end{theorem}

\begin{proof}
The proof will be a modif\/ication of the proof of
Proposition~\ref{prop2}. Note f\/irst that the Killing form on
$\gl{g}$ induces an $\Ad$-invariant inner product on $L\gl{g}$
given by the formula
\[
\leb \xi (x) , \eta (x) \rib = \int _{S^1} \langle \xi (x), \eta
(x) \rangle \, \dd x .
\]
By the symbol $\leb - , - \rib$ we shall also denote the pairing
between the elements of $L\gl{g}$ and those of~$L\gl{g}^*$, as
well as the induced inner product on the dual $L\gl{g}^*$. With
this notation the directional derivative of the Hamiltonian
$H_{mb}$ in the direction $(\delta_b, \delta_{ct})$ is given by
\[
\leb \dd H_{mb}, (\delta _b, \delta _{ct}) \rib = - \leb [\sigma,
\Ad_g(\tau)]^{\flat}, \delta_b \rib + \leb \delta _{ct},
p_g^{\sharp} \rib,
\]
where the maps $\sharp$ and $\flat$ are def\/ined in the same way
as before.

The formula (\ref{ctg}) from the Proposition \ref{prop2} is valid
for every Lie group, hence also for the loop group $LG$. Therefore
we get the following expression for the form $\omega _m + c
\omega_m $:
\begin{gather*}
(\omega_c + c   \omega_m)_{(g, p_g)} ((X_b, X_{ct}), (Y_b, Y_{ct})) \\
\qquad{}  =  - \leb X_{ct}, Y_b \rib + \leb Y_{ct}, X_b \rib    +
\leb p_g, [X_b, Y_b] \rib - c   \leb (X_b)_x, Y_b \rib .
 \end{gather*}
 The above two formulae now give us the following relation for the
 Hamiltonian vector f\/ield $X_{H_{mb}} = (X_b, X_{ct})$:
 \begin{gather*}
 (\omega _c + c   \omega_m)_{(g, p_g)} ((X_b, X_{ct}), ( \delta _b, \delta _{ct})) \\
 \qquad{} =
 - \leb X_{ct}, \delta _b \rib + \leb  \delta _{ct}, X_b \rib +
  \leb p_g, [X_b, \delta_b ] \rib -
  c   \leb (X_b)_x, \delta _b \rib  \\
\qquad {}= \leb - X_{ct} - c   (X_b)_x^{\flat} - \{X_b, p_g\},
\delta _b \rib +
 \leb \delta _{ct}, X_b \rib .
\end{gather*}
Since the vectors $\delta _b$ and $\delta_{ct}$ are linearly
independent, we get two equations for the components of the
Hamiltonian vector f\/ield:
 \begin{gather}
 X_b = p_g^{\sharp} , \qquad   X_{ct} +
 c  (X_b)_x^{\flat} =  [\sigma, \Ad_g(\tau)]^{\flat}.
 \label{loopcanonical}
\end{gather}
 Let $\gamma(t, x) = (g(t, x), p_g(t, x)) \colon \rf \to T^*LG$
 be an integral curve of the f\/ield $X_{H_{mb}}$ given in the right trivialization
 $T^*LG \cong LG \times L\gl{g}$. Then we have
 $\dot{\gamma} = (g_tg^{-1}, (p_g)_t)$ and thus from
 (\ref{loopcanonical}) f\/inally
 \begin{gather*}
 \big(g_tg^{-1}\big)_t + c   \big(g_tg^{-1}\big)_x = [\sigma, \Ad_g(\tau)].\tag*{\qed}
 \end{gather*}\renewcommand{\qed}{}
\end{proof}

\section{A family of conservation laws}\label{symmetries}

In this section we shall construct a family of conservation laws
for the generalized Maxwell--Bloch system. This family will
correspond to a certain group of symmetries of the Maxwell--Bloch
system. The symmetries form a loop group. Hence the corresponding
family of conservation laws  will consist of inf\/initely many
functionally independent elements.

Let, for the sake of simplicity, the loop $\tau(x)$ in the
equation
\[
\big(g_tg^{-1}\big)_t + c   \big(g_t g^{-1}\big)_x = [ \sigma,
\Ad_g (\tau)]
\]
be constant, $\tau (x) \equiv \tau$. Denote by $\gl{t}_{\tau}$ the
maximal toroidal subalgebra in $\gl{g}$ which contains the element
$\tau$. Let $T_{\tau} = \exp{\gl{t}_{\tau}}$ be the corresponding
maximal torus in $G$. We notice immediately that our equation is
invariant with respect to the right  action
\[
\rho_{r(x)}(g(t, x)) = g(t, x) \cdot r(x), \qquad r(x) \colon S^1
\to T_{\tau}
\]
of the loop group $LT_{\tau} \subset LG$. Our conserved quantities
will correspond to the $LT_{\tau}$-symmetry in the sense of
Noether's theorem.

Equivalently, the generalized Maxwell--Bloch Hamiltonian system
$(T^*LG, \omega _c + c   \omega _m, H_{mb})$ is invariant with
respect to the canonical lifting $\varrho$  of the action $\rho$
onto the cotangent bundle $T^*LG$. It is easily checked that, in
the trivialization of $T^*LG$ by the right translation, the action
$\varrho \colon LT_{\tau} \times T^*LG \to T^*LG $ is given by
\[
\varrho _{r(x)}\bigl(g(x),   p_g(x)\bigr) = \bigl( g(x)\cdot r(x),
p_g(x)\bigr).
\]
This is an action by symplectomorphisms. Indeed, the lifting to
the cotangent bundle of an action on the base space is always
symplectic. But this right action is symplectic also with respect
to the form $\omega_m$, due to the fact that $\omega_m$ is
right-invariant. Thus the action $\varrho$ is Hamiltonian, since
$T^*LG$ is a simply connected space. Therefore this action has the
moment map. Our   conservation laws will be the components of the
moment map corresponding to the action~$\varrho$. We shall prove
the following theorem.

\begin{theorem} Let $(T^*LG, \omega_c + c   \omega_m, H_{mb})$ be the
generalized Maxwell--Bloch system. Let $ \xi (x) \in
L\gl{t}_{\tau} $ be an arbitrary loop in $\gl{t}_{\tau}$. Then the
function $F_{\xi} \colon T^*LG \to \rf$, which in the right
trivialization of $T^*LG$ is defined by the formula
\[
F_{\xi} \bigl( g(x),    p_g(x)\bigr) = \int _{S^1} \left( p_g(
\Ad_g(\xi))  - \frac{c}{2} \langle g_xg^{-1} , \Ad_g(\xi) \rangle
\right)   \dd x,
\]
is a first integral of this system. In other words, let
\[
g(t, x) \colon \rf \times S^1 \to G
\]
be a solution of the Maxwell--Bloch equation \eqref{MBN1}. Then
for every loop $\xi(x) \colon S^1 \to \gl{t}_{\tau}$, the quantity
\[
G_{\xi}\bigl( g(t, x)\bigr) = \int _{S^1} \langle g_t g^{-1} -
\frac{c}{2}   g_x g^{-1},
  \Ad_{g}(\xi)  \rangle   \dd x
\]
is constant with respect to  time $t$.
\end{theorem}

\begin{proof}
The second part of the theorem follows immediately from the
f\/irst part. Let $\gamma (t) = (g(t), p_g(t)) \colon \rf \to
T^*LG$ be a solution of the Hamiltonian system $(T^*LG,  \omega_c
+ c   \omega_m, H_{mb})$, and let $g(t, x) \colon \rf \times S^1
\to G$ be the map given by the curve $g(t) = \pi(\gamma (t))\colon
\rf \to LG$, where $\pi \colon T^*LG \to LG$ is the natural
projection. Then
\[
g(t, x) \colon \rf \times S^1 \to G
\]
is a solution of the Maxwell--Bloch equation (\ref{MBN1}). Again
we shall use the trivialization $T^*LG \cong LG \times \gl{g}^*$
by the right translations. Denote by
\[
pr_2 \colon T^*LG \cong LG \times \gl{g}^* \to L\gl{g}^*
\]
the projection on the second component. Then the Killing form on
$G$ yields the  identif\/ication
\[
pr_2 (\gamma (t) ) = p_g(t) = \langle g_t(t, x)g^{-1}(t, x)   ,
- \rangle.
\]
This clearly shows that
\[
F_{\xi}\bigl(\gamma(t)\bigr) = \int _{S^1} \langle g_tg^{-1} -
\frac{c}{2}   g_xg^{-1} ,   \Ad_g(\xi) \rangle
 \dd x =  G_{\xi}\bigl(g(t, x)\bigr)
\]
for every $t$.

We shall now show that $F_{\xi}$ are integrals of the
Maxwell--Bloch Hamiltonian system. We have already mentioned that
our integrals are the components of the moment map corresponding
to the action $\varrho$ of $TL_{\tau}$ on the Maxwell--Bloch
system.

Let $\xi (x) \in L\gl{t}_{\tau}$ be an arbitrary element in the
Lie algebra $L\gl{t}_{\tau}$. Denote by $\widetilde{\xi}$ the
inf\/initesimal action of $\xi$ on $T^*LG$. By def\/inition, the
$\xi$-component $H_{\xi}$ of the moment map $ \mu \colon T^*LG \to
(L\gl{t}_{\tau})^* $ is given by the formula
\[
\mu(g, p_g) (\xi) = H_{\xi}(g, p_g),
\]
where $H_{\xi} \colon T^*LG \to \rf$ is the function whose
Hamiltonian vector f\/ield is the inf\/initesimal
action~$\lel{\xi}$. This means that
\[
(\dd H_{\xi})_{(g, p_g)}(\delta_b, \delta_{ct}) = (\omega_c + c
\omega_m) (\lel{\xi}, (\delta_b, \delta_{ct}))  = \omega_c
(\lel{\xi}, (\delta_b, \delta_{ct})) +
   c   \omega_m (\lel{\xi}, (\delta_b, \delta_{ct}))
\]
for every tangent vector $(\delta_b, \delta_{ct}) \in T_{(g,
p_g)}(T^*LG)$. Let the functions $H_{\xi}^c, H_{\xi}^m \colon
T^*LG \to \rf$ be given by the formulae
\[
(\dd H_{\xi}^c)_{(g, p_g)}(\delta_b, \delta_{ct}) = (\omega_c
)_{(g, p_g)}(\lel{\xi}, (\delta_b, \delta_{ct})), \qquad (\dd
H_{\xi}^m)_{(g, p_g)}(\delta_b, \delta_{ct}) = (\omega_m)_{(g,
p_g)}(\lel{\xi}, (\delta_b, \delta_{ct})).
\]
Then
\begin{gather}
H_{\xi} = H_{\xi}^c + c   H_{\xi}^m. \label{one}
\end{gather}
Here we ignore the non-relevant indeterminate additive constant,
and we shall continue to do so below. For the canonical form we
have $\omega_c =  \dd \theta$, where $\theta $ is the tautological
1-form on $T^*LG$. Thus we have
\[
(\dd H_{\xi}^c)_{(g, p_g)}(\delta_b, \delta_{ct}) = (\dd
\theta)_{(g, p_g)} (\lel{\xi}, (\delta_b, \delta_{ct})), \qquad
{\rm therefore} \qquad H_{\xi}(g, p_g) = \theta_{(g,
p_g)}(\lel{\xi}).
\]
The expression of the inf\/initesimal action in the right
trivialization is given by
\[
\lel{\xi}(g, p_g) =  \frac{\dd}{\dd s} \Big|_{s = 0} \varrho
_{h(s)}(g, p_g) =  (\Ad_g(\xi), p_g).
\]
Here $h(s) \colon (- \epsilon, \epsilon) \to LT_{\tau}$ is a path,
such that $h(0) = e$, and $\frac{{\rm d }}{{\rm d} s}\big|_{s = 0}
h(s) = \xi \in L\gl{t}_{\tau}$. From the def\/inition of the
tautological 1-form we now have
\begin{gather}
H_{\xi}^c (g, p_g) = p_g(\lel{\xi}) =  \int _{S^1}
p_g(\Ad_g(\xi))\,   \dd x, \label{two}
\end{gather}
which gives the f\/irst summand in (\ref{one}).

We claim the the magnetic component $H_{\xi}^m$ is given by the
formula
\begin{gather}
H_{\xi}^m(g, p_g) =  - \frac{1}{2}   \leb g_xg^{-1}, \Ad_g(\xi)
\rib = - \frac{1}{2}   \int _{S^1} \langle g_x g^{-1}, \Ad_g(\xi)
\rangle   \dd x. \label{three}
\end{gather}
We will calculate the derivative of $H_{\xi}^m$ at $(g, p_g)$ in
the direction $(\delta_b, \delta_{ct}) \in T_{(g, p_g)}(T^*LG)$.
Let $s \mapsto (g(s), p_g(s))$ be a path, such that $(g(0),
p_g(0)) = (g, p_g)$ and $\frac{{\rm d} }{{\rm d} s}\big|_{s = 0}
(g(s), p_g(s) ) = (\delta_b, \delta_{ct})$ in the right
trivialization. Derivation gives
\begin{gather*}
\dd (H_{\xi}^m)_{(g, p_g)}(\delta_b, \delta_{ct})   =
\frac{\dd }{\dd s}  \Big|_{s = 0} \int _{S^1} \langle g(s)_xg(s)^{-1}, \Ad_{g(s)}(\xi) \rangle \\
\phantom{\dd (H_{\xi}^m)_{(g, p_g)}(\delta_b, \delta_{ct}) }{} =
\int _{S^1} \bigl( \langle (\delta _b)_x, \Ad_g(\xi) \rangle +
\langle g_xg^{-1},
[\delta _b, \Ad_g(\xi) ] \rangle \bigr) \,  \dd x \\
\phantom{\dd (H_{\xi}^m)_{(g, p_g)}(\delta_b, \delta_{ct}) }{} =
\int _{S^1} \bigl(-   \langle \delta _b, \bigl(\Ad_g(\xi)\bigr)_x
\rangle -
\langle \delta _b, [ g_x g^{-1}, \Ad_g(\xi)] \rangle  \bigr)  \, \dd x\\
\phantom{\dd (H_{\xi}^m)_{(g, p_g)}(\delta_b, \delta_{ct}) }{}
 =  - 2 \int _{S^1} \langle \delta _b, [ g_x g^{-1}, \Ad _g(\xi) ] \rangle\,   \dd x .
\end{gather*}
On the other hand we have
\[
(\omega_m)_{(g, p_g)}\bigl(\lel{\xi}, (\delta _b, \delta
_{ct})\bigr) = \int _{S^1} \langle (\Ad _g(\xi))_x, \delta _b
\rangle   \, \dd x = \int _{S^1} \langle \delta _b, [ g_xg^{-1},
\Ad_g(\xi) ] \rangle  \,  \dd x,
\]
that proves (\ref{two}).

If we now put together the formulae (\ref{one}), (\ref{two}) and
(\ref{three}), we f\/inally get
\[
H_{\xi}(g, p_g) = \int _{S^1} \left( p_g(\Ad_g(\xi)) - \frac{c}{2}
\langle g_x g^{-1} , \Ad_g(\xi) \rangle \right)   \dd x
\]
that proves the theorem.
\end{proof}

Our conservation laws can be expressed in a more compact and
suggestive form. Let us choose an arbitrary point $x_0 \in S^1$
and let $\varphi_n(x) \colon S^1 \to \rf$ be a sequence of
functions which approximates the Dirac delta function
$\delta_{x_0}$ on $S^1$. For every function $f \colon S^1 \to \rf$
we then have
\[
\lim _{n \to \infty} \int _{S^1} f(x) \cdot \varphi _n(x) \,  \dd
x = \int_{S^1} f(x) \cdot \delta_{x_0} (x)  \, \dd x = f(x_0).
\]
Let  $\hat{\xi} \in \gl{t}_{\tau}$ be an arbitrary element and let
$\xi_n(x) \in L\gl{t}_{\tau}$ be elements in the loop algebra
def\/ined by $\xi _n(x) = \varphi_n(x) \cdot \hat{\xi}$. Then it
follows from the above theorem that for every $n \in \mathbb{N}$
we have
\[
\int _{S^1} \langle g_tg^{-1} - \frac{c}{2} g_xg^{-1} , \Ad_g(\xi
_n) \rangle    \,\dd x = \int _{S^1} \left( \langle g_tg^{-1} -
\frac{c}{2} g_xg^{-1} , \Ad_g(\hat{\xi}) \rangle (x)  \cdot
\varphi _n (x) \right)    \dd x = A_n.
\]
The sequence $\{A_n\}_{n \in \mathbb{N}}$ of constants is clearly
convergent and its limit is
\[
\langle g_tg^{-1} - \frac{c}{2}   g_xg^{-1} , \Ad _g(\hat{\xi})
\rangle (x_0) = \int _{S^1} \left( \langle g_tg^{-1} - \frac{c}{2}
g_xg^{-1} , \Ad _g(\hat{\xi}) \rangle (x) \cdot \delta_{x_0} (x)
\right)    \dd x.
\]
This proves the following corollary.
\begin{corollary}
For every element $\hat{\xi}$ in the maximal Abelian subalgebra
$\gl{t}_{\tau} \subset \gl{g}$ and for every point $x_0 \in S^1$,
the quantity
\[
\chi _{x_0}\bigl( g(t, x) \bigr) = \langle g_t g^{-1} -
\frac{c}{2}   g_xg^{-1}, \Ad_g(\hat{\xi}) \rangle (x_0)
\]
is constant along every solution $g(t, x) \colon \rf \times S^1
\to G$ of the generalized Maxwell--Bloch equation.

From the $\Ad$-invariance of the Killing form $\langle - , - \rangle$ we see that
for every $\hat{\xi} \in \gl{t}_{\tau}$ the quantity
\[
\chi  _{x_0} = \langle g^{-1} g_t - \frac{c}{2}   g^{-1}g_x,
\hat{\xi} \rangle (x_0)
\]
is constant. To put it more briefly, let $p_{\gl{t}_{\tau}} \colon
\gl{g} \to \gl{t}_{\tau}$ be the orthogonal projection. Then for
every solution $g(t, x) \colon \rf \times S^1 \to G$ and for every
point $x_0 \in S^1$ the projection
\[
p_{\gl{t}_{\tau}}\left( \left(g^{-1} g_t - \frac{c}{2}
g^{-1}g_x\right)(x_0) \right)
\]
is a constant element in $\gl{t}_{\tau}$.
\end{corollary}

Consider again the Hamiltonian function $ H_{mb}(g, p_g) $ of the
Maxwell--Bloch dynamical system $(T^*LG, \omega_c + c   \omega_m,
H_{mb})$. We see that the left action $\rho^l_s(x) (g(x)) = s(x)
\cdot g(x)$ of the subgroup $LT_{\sigma} \subset LG$ preserves the
Hamiltonian  $H_{mb}$. Here $T_{\sigma}$ is the maximal torus in
$G$ whose Lie algebra~$\gl{t}_{\sigma}$ contains the
element~$\sigma$. But the group $LT_{\sigma}$ does not preserve
the magnetic part~$\omega_m$ of the symplectic structure $\omega_c
+ c   \omega_m $. Recall that the magnetic part is right
invariant, but it is not left invariant -- not even with respect
to the action of $LT_{\sigma}$. It is easily seen that only the
torus $T_{\sigma} \subset LT_{\sigma}$ (containing the constant
loops) preserves the symplectic structure, and hence the whole
Hamiltonian system. Calculations, similar to those above, only
simpler, give us the proof of the following proposition.
\begin{proposition}
Let $\eta \in \gl{t}_{\sigma}$ be an arbitrary element. Then for
every solution $g(t, x) \colon ~\rf \times S^1 \to G$ the quantity
\[
J_{\eta} \bigl( g(t, x) \bigr) = \int _{S^1} \langle g_t g^{-1},
\eta \rangle  \, \dd x
\]
is a constant. This means that the element
\[
P_{\gl{t}_{\sigma}}\bigl( g(t, x) \bigr) = p_{\gl{t}_{\sigma}}
\left( \int _{S^1} g_tg^{-1}    \dd x  \right)
\]
of $\gl{t}_{\sigma}$ is constant along every solution. Here
$p_{\gl{t}_\sigma } \colon \gl{g} \to \gl{t}_{\sigma}$  is again
the orthogonal projection.
\end{proposition}

It is well known that the Maxwell--Bloch equations are integrable.
This is also true for the generalized Maxwell--Bloch equations. In
particular, they satisfy the zero-curvature condition
\[
V_t - U_x + [U, V] = 0
\]
for the Lax pair
\[
U   =  - \big( - z \sigma + g_t g^{-1}\big)  \qquad {\rm and}
\qquad V   =   - z \sigma + g_tg^{-1} - \frac{1}{z} \Ad _g (\tau).
\]
The proof is a matter of trivial checking.  We have to note that
the conserved quantities constructed above do {\it not} comprise a
complete system of f\/irst integrals. But they can be used to
reduce the system. An interested reader can f\/ind various results
concerning the integrability of the Maxwell--Bloch equations in
the references \cite{L1, L2,CGB,GZM} etc.

As we mentioned above, the  conserved quantities associated with
the action of $LT_{\tau}$ can be used to construct symplectic
reductions of the system $(T^*LG, \omega_c + c   \omega _m,
H_{mb})$. One can see that the nature of the symplectic reductions
$\mu^{-1}(\alpha)/LT_{\tau}$ depends quite heavily on the choice
of the level $\alpha \in L\gl{t}_{\tau}$ of the moment map. We
intend to study some cases in another paper.

For the case of the original Maxwell--Bloch equation, these
symplectic quotients can be thought of as inf\/inite-dimensional
analogues of the following situation. Let $(T^*SU(2), \omega _c,
H)$,
\[
H(g, p_g) = \frac{1}{2} \| p_g\|^2 + \langle \sigma, \Ad _g(\tau)
\rangle,
\]
be the Hamiltonian system of the C.~Neumann oscillator on $S^3$
with two circular symmetries, or equivalently, the Maxwell--Bloch
system for the solutions which are constant with respect to the
spatial variable.  Then the right action of the maximal torus
$T_{\tau} = \exp{\gl{t}_{\tau}}$ on our system is Hamiltonian. Let
$\mu \colon T^*G \to \gl{t}_{\tau}$ be the corresponding moment
map. The  quotient $(\mu^{-1}(a) /T_{\tau}, \omega_{sq}, H_{sq})$
is the system $(T^* S^2, \omega_c + a   \omega _m, H_{sq})$, where
the induced Hamiltonian is given by
\[
H_{sq}(q, p_q) = \frac{1}{2}\| p_q\|^2 + \langle \sigma, q \rangle
\]
and $q = \Ad_g(\tau) \in \orb = S^2 \subset \su (2)$. The magnetic
perturbation $\omega _m$ of the symplectic form is given by
\[
(\lel{\omega}_m)_{(q, p_q)}(X_1, X_2) = p_q ( [ \kappa _1, \kappa
_2]), \qquad X_i = [\kappa _i, q] \in T_q \orb = T_q S^2.
\]
The reduced system describes the charged spherical pendulum which
moves in the magnetic f\/ield of a Dirac monopole situated in the
centre of the sphere $S^2$. More details and the proof can be
found for example in \cite{Sa1,Sa2}. The quotient construction is
easily generalized to the cases when $SU(2)$ is replaced by a
compact semi-simple $G$. The sphere is then replaced by the
general co-adjoint orbit $\orb$  of $\tau$ and the magnetic form
$\omega _m$ is just the well-known Kostant--Kirillov symplectic
form   which is a part of the K\"ahler structure on~$\orb$.

\section {Concluding remarks}

We have described a new Hamiltonian structure of the
Maxwell--Bloch equations. The key step in the construction is the
observation that the Maxwell--Bloch equations can be considered as
the equation of motion of a continuous chain of C. Neumann
oscillators on the three-dimensional sphere. These oscillators
interact through forces of the magnetic type. Our Hamiltonian
structure is  derived from the well-known Hamiltonian structure of
the C. Neumann oscillator. What has to be added in the case of the
Maxwell--Bloch equations is the component which accounts for the
magnetic-type interactions among the oscillators in the chain.
This is achieved by means of the perturbation of the canonical
symplectic structure by a topologically non-trivial magnetic term.
This term is studied in more detail in \cite{Sa}. Among other
things, we show in \cite{Sa} that in the Lagrangian formulation
this term gives rise to essentially the same topologically
non-trivial term as the one appearing in the Wess--Zumino--Witten
theory. In this paper we have concentrated on f\/inding a family
of conservation laws of the Maxwell--Bloch equations. The task was
facilitated exactly by the use of our particular Hamiltonian
structure.

It is important to note that the Maxwell--Bloch equations might
very well be endowed with additional Hamiltonian structures,
dif\/ferent form the one described here. One such structure is
given by Holm and Kova\v ci\v c in \cite{HoKo}.  It can be easily
seen that the two structures are inequivalent. The symplectic
structure presented in \cite{HoKo} does not include derivatives
with respect to the spatial variable, while ours does. The
existence of these two Hamiltonian structures suggests the
possibility  of the existence of a bi-Hamiltonian structure on the
Maxwell--Bloch system. We intend to address this issue in another
paper. Another interesting avenue to the study of  Hamiltonian
structures was suggested to the author by one of the referees.
He/she points out the important fact that the reduced
Maxwell--Bloch equations can re rewritten in the form of a
nonlinear von Neumann equation. This is shown in \cite{NK}. A
treatment of the nonlinear von Neumann-type equations and of their
importance to quantum physics can be found in \cite{NCKL}. Some of
the nonlinear von Neumann equations indeed have the Hamiltonian
formulations of dif\/ferent sorts. An interesting example pointed
out by the referee is the generalized Hamiltonian structure of
Nambu, described in \cite{Na}.

Let us brief\/ly outline a somewhat more traditional way of
describing the Hamiltonian structure of the {\it reduced}
Maxwell--Bloch equations.  The reduced system equations considered
in \cite{NK} (and in other sources) is the system of three
ordinary dif\/ferential equations
\begin{alignat}{3}
& \dot{u}_1  =  - \Delta  u_2,  \qquad && \Delta   = {\rm  const},\nonumber\\
& \dot{u}_2   =  \Delta u_1 + k {\cal E} u_3,  \qquad &&  k  =  {\rm const},\nonumber\\
& \dot{u}_3   =  - k {\cal E} u_2, \qquad &&  {\cal E}  =(2/k
\tau) \, {\rm sech}(t/\tau).& \label{reduced}
\end{alignat}
Since the variables $u_1$, $u_2$ and $u_3$ are related by the
constraint $u_1^2 + u_2^2 + u_3^2 = 1$, the phase space of this
system is the two-dimensional sphere $S^2$. We can view $S^2$ as a
co-adjoint orbit of a non-zero element in the Lie algebra $\su (
2)$. If we introduce $\su (2)$-valued maps $\varrho \colon  S^2
\to \su (2)$ and $\Omega \colon S^2 \times \rf \to \su (2)$
\[
\varrho =  \begin{pmatrix} i u_3 & u_1 + i u_2  \\
                          - u_1 + i u_2& - i u_3 \end{pmatrix},
      \qquad
\Omega (t) = \frac{1}{2}\begin{pmatrix} i \Delta & \kappa {\cal E}(t)  \\
                          - \kappa {\cal E}(t)& - i \Delta \end{pmatrix},
\]
then the system (\ref{reduced}) can be written in the form
\begin{gather}
\dot{\varrho} = [\Omega, \varrho]. \label{su2red}
\end{gather}
Let us recall the Kostant--Kirillov symplectic form on the
arbitrary co-adjoint orbit ${\cal O} \subset \gl{g}$ given by the
formula
\[
(\omega_{kk} )_m(X, Y)= m ([a, b]), \qquad  X =[m, a], \    Y = [m
, b] \in T_m{\cal O}.
\]
In the case of $\gl{g} = \su (2)$, we can identify $\su(2)$ and
the dual space $\su (2)^*$ via the Killing form $\langle \alpha,
\beta \rangle = - \frac{1}{2}\,{\rm Tr} (\alpha \cdot \beta)$.
Thus, on $S^2 \subset \su (2)$ we get the expression
\[
(\omega _{kk})_{\varrho}(X, Y) = \langle \varrho, [a, b]\rangle
 = - \frac{1}{2} \,{\rm Tr}\,(\varrho \cdot [a, b]), \qquad
 X =[\varrho, a], \   Y = [\varrho , b] \in T_{\varrho} S^2.
\]
Let us now consider the function $H \colon S^2 \times \rf \to \rf$
given by the formula
\begin{gather}
H(\varrho, t) = \langle \varrho, \Omega (t)\rangle.
\label{thamiltonian}
\end{gather}
A short calculation shows that the reduced Maxwell--Bloch
equations (\ref{su2red}) are the equation of motion for the
Hamiltonian system $(T^*S^2, \omega_{kk}, H)$ with the
time-dependent Hamiltonian given by the formula
(\ref{thamiltonian}).

We conclude this remark by observing that,  in one way or another,
the reduced Maxwell--Bloch equations lead to Hamiltonian
structures which are dif\/ferent from the one considered in our
paper. The main reason for this probably lies in the fact that in
the case of the reduced equations, the electric f\/ield ${\cal E}$
is not an unknown, but it is assumed to be a known function of
time, for example ${\cal E}(t) = (2/k \tau) \ {\rm sech}(t/\tau)$.

In this paper, we studied the Maxwell--Bloch equations in their
slowly varying envelopes approximation and in the sharp line
limit, that is, without the inhomogeneous broadening. 
When the inhomogeneous broadening is taken into account, the resulting equations
preserve the families of symmetries described in this paper. These symmetries can
then be used for reduction and the resulting reduced equations would be analogous
to those described above. However, the equations with broadening do not have
a Hamiltonian structure directly analogous to the one described in this paper. For
any Hamiltonian structure of the equations with broadening, the symplectic
structure would have to be substantially dif\/ferent from ours. 
At this moment the author does not know, whether it would
also carry over to the case without the assumption of the slowly
varying envelopes.

\subsection*{Acknowledgements}

I would like to thank professors Pavel Winternitz, Gregor Kova\v ci\v c, 
and Ji\v r\' i Patera for  interesting and stimulating
discussions. The research for this paper was supported in part by
the research programme Analysis and Geometry P1-0291, Republic of
Slovenia. A part of the research was done at the Centre de
Recherches Math\'ematiques, Montreal, Canada. The  hospitality of
CRM and especially of professor Ji\v r\' i Patera is gratefully
acknowledged.

\LastPageEnding

\end{document}